\documentclass[10pt,journal,twocolumn]{IEEEtran}

\makeatletter
\let\NAT@parse\undefined
\makeatother

\usepackage[numbers]{natbib}

\usepackage[cmex10]{amsmath}
\usepackage{amssymb}
\usepackage{xy}
\usepackage[vflt]{floatflt}
\usepackage[dvips]{graphicx}
\DeclareGraphicsExtensions{.eps}

\usepackage{hyperref}

\newcommand{\N}{\mathbb{N}}
\newcommand{\Q}{\mathbb{Q}}

\newfont{\hiera}{cmsy10 scaled 2488} 
\newfont{\hierb}{cmsy10 scaled 1728}
\newfont{\hierc}{cmsy10 scaled 1200}

\newcommand{\Bigast}{
\mathop{\vphantom{\sum}\lower2.5pt\hbox{\hiera\char3}}}%

\newcommand{\Bigtimes}{
\mathop{\vphantom{\sum}\lower2.5pt\hbox{\hiera\char2}}}%

\def\BibTeX{{\rm B\kern-.05em{\sc i\kern-.025em b}\kern-.08em
    T\kern-.1667em\lower.7ex\hbox{E}\kern-.125emX}}

\newtheorem{theorem}{Theorem}

\newtheorem{example}[theorem]{Example}

\begin{document}
\title{\Large ``E pluribus unum''\\
or\\
How to Derive Single-equation Descriptions for Output-quantities in Nonlinear Circuits using Differential Algebra\\
{\large (2008 Re-Release)}
}

\author{Eberhard~H.-A.~Gerbracht
\thanks{This article first appeared in: Proceedings of the 7th International Workshop on Symbolic Methods and Applications to Circuit Design, SMACD 2002, Sinaia, Romania, October 10-11, 2002. Bukarest 2002, pp.~65--70. Due to the low distribution of these proceedings, the author has decided to make the article available to a wider audience through the arXiv.}
\thanks{At the time of origin of this paper the author was with the Institut f\"ur Netz\-werk\-theorie und Schaltungstechnik, Technische Universit\"at Braunschweig, D-38106 Braunschweig, Germany.}
\thanks{His current (April 17th, 2008) address is Bismarckstra\ss e 20, D-38518 Gifhorn, Germany. Current e-mail: \tt{e.gerbracht@web.de}}}
 
\maketitle
\thispagestyle{empty}
\pagestyle{empty}

\begin{center}
{\sl Dedicated to the memories of}

{\sl Prof.\ Dr.-Ing.\ Ernst-Helmut Horneber (1946--2001)} 

{\sl and}

{\sl Prof. Giuseppa Carr\`a Ferro (1952--2007)}
\end{center}

\bigskip

\begin{abstract}
 In this paper we describe by a number of examples how to deduce one single characterizing higher order differential equation for output quantities of an analog circuit.

  In the linear case, we apply basic "symbolic" methods from linear algebra to the system of differential equations which is used to model the analog circuit.
For nonlinear circuits and their corresponding nonlinear differential
equations, we show how to employ computer algebra tools implemented in Maple,
which are based on differential algebra.
\end{abstract}

{\small
\smallskip

}


\section{Introduction}

Usually the input-output response of a linear time-invariant circuit is described in the frequency domain by its transfer function, i.e.\ a single rational function. This translates directly into a linear differential equation with constant coefficients in the time domain. The advantage of this approach is, that in any guise, only one (differential) equation is needed to completely describe the quantity a designer is interested in.

This method fails miserably, when a transformation from the time to the frequency domain is not possible, e.g., when nonlinear circuits have to be examined, which lead to systems of nonlinear differential equations. Even though most of the times, 
these can be given in symbolic terms, any single quantity in the circuit usually is described by a ``waveform'', which results from assigning a numerical value to each symbol, followed by a numerical calculation using computer simulation. The advantage of having only {\sl one } describing equation seems to have been irrevocably lost, when nonlinear circuits are considered.  Thus, up until now, nonlinear circuits seemed nearly inaccessible to most symbolic approaches.

This problem is not a new one, and it is not limited to the area of analog circuits alone. Several years ago, researchers in nonlinear control theory have proposed to use constructive methods from differential algebra to tackle their problems. 
At the same time -- and inspired in part by this proposal -- mathematicians started to implement algorithms from differential algebra, which had already been formulated in the 1950s. 
These programmes became part of the {\tt MAPLE} computer algebra system.

At the SMACD-meeting in 1998 G.\ Carr\`a Ferro\footnote{Note added in 2008: Between the time of this article's first publication and the publication of the electronic version, Giuseppa Carr\`a Ferro passed away on March 22nd, 2007. Thus  a dedication to her memory was added to this electronic version.} gave examples of how to transform systems of nonlinear differential equations containing certain transcendental functions, that arise from analog circuits, into systems of nonlinear ''algebraic'' differential equations \citep{Carra_SMACD}, and thus brought the area of constructive differential algebra in contact with the area of symbolic circuit analysis and design. In this paper we will take this approach several steps further and show, how single equations for any quantity in a circuit can be derived from these systems. We will be able to give a new and easy algorithm for linear circuits, which works in the time domain, and uses only differentiation and Gaussian elimination. For nonlinear circuits we will resort to the algorithms already implemented in {\tt MAPLE}. We will comment on how to apply them and produce several examples.


\section{Linear Circuits and linear Systems of Differential Equations}

To get a first flavour of things, we start our discussion with linear circuits and their corresponding systems of linear differential equations. But, instead of working in the frequency domain, using the Laplace-transform to deduce the transfer function for a sought-after quantity and thus its characterizing differential equation, we will remain in the time domain.

\subsection{Linear State Equations}

In this section we will present a new algorithm, that, starting from a set of state equations, only uses repeated differentiation and, finally, Gaussian elimination, to compute the single differential equation for any given quantity. We will restrict our presentation to just three state variables $x_1, x_2, x_3,$ but it is easy to extend the method, shown below, to any number of state variables.

So, let us suppose, that a given linear circuit can be described by a set of linear state equations. We ask for one single differential equation describing, without loss of generality, the state variable $x_1.$ Again it is easy to handle other state variables or any output variable $y,$ which is a linear combination of state variables and inputs, in an analogous manner.

Let the system be given by
\begin{equation}
\begin{split}
\label{StateEq}
\dot x_1(t) & =  a_{11}\cdot x_1(t) + a_{12}\cdot x_2(t) + a_{13}\cdot x_3(t) + e_1(t),\\
\dot x_2(t) & =  a_{21}\cdot x_1(t) + a_{22}\cdot x_2(t) + a_{23}\cdot x_3(t) + e_2(t),\\
\dot x_3(t) & =  a_{31}\cdot x_1(t) + a_{32}\cdot x_2(t) + a_{33}\cdot x_3(t) + e_3(t),
\end{split}
\end{equation}

\noindent
where $x_1,x_2,x_3$ denote the state variables and $e_1,e_2,e_3$ represent linear combinations of the inputs (and eventually their derivatives).


Clearly, if $x_1,x_2,x_3$ satisfy (\ref{StateEq}), their derivatives\footnote{In the sequel wherever necessary, we will assume that all expressions are differentiable over a suitable extension field of the real numbers.} will satisfy
\begin{equation}
\label{StateEqDer1}
\begin{split}
\ddot x_1(t) & =  a_{11}\cdot \dot x_1(t) + a_{12}\cdot \dot x_2(t) + a_{13}\cdot \dot x_3(t) + \dot e_1(t),\\
\ddot x_2(t) & =  a_{21}\cdot \dot x_1(t) + a_{22}\cdot \dot x_2(t) + a_{23}\cdot \dot x_3(t) + \dot e_2(t),\\
\ddot x_3(t) & =  a_{31}\cdot \dot x_1(t) + a_{32}\cdot \dot x_2(t) + a_{33}\cdot \dot x_3(t) + \dot e_3(t).
\end{split}
\end{equation}

\noindent
If $n>3$ state variables are given, we have to repeat the above procedure $n-1$ times.
It is a well known observation (which will be shown as a byproduct of our algorithm), that $n$ linear first order state equations lead to one $n$th order differential equation for any single quantity. Thus, in our example, we need one further equation for the third derivative $\dddot x_1$ of $x_1$, which we get by differentiating once again the first equation of (\ref{StateEqDer1}).

\begin{equation}
\label{StateEqFinalDr}
\dddot x_1(t) = a_{11}\cdot \ddot x_1(t) + a_{12}\cdot \ddot x_2(t) + a_{13}\cdot \ddot x_3(t) + \ddot e_1(t).
\end{equation}

\noindent
Next, we write down all of the above equations into one system, where the vector of variables is given by all the derivatives of all state variables, that have been produced by the above procedure. 
\begin{center}
{\includegraphics[width=\linewidth,clip,keepaspectratio]{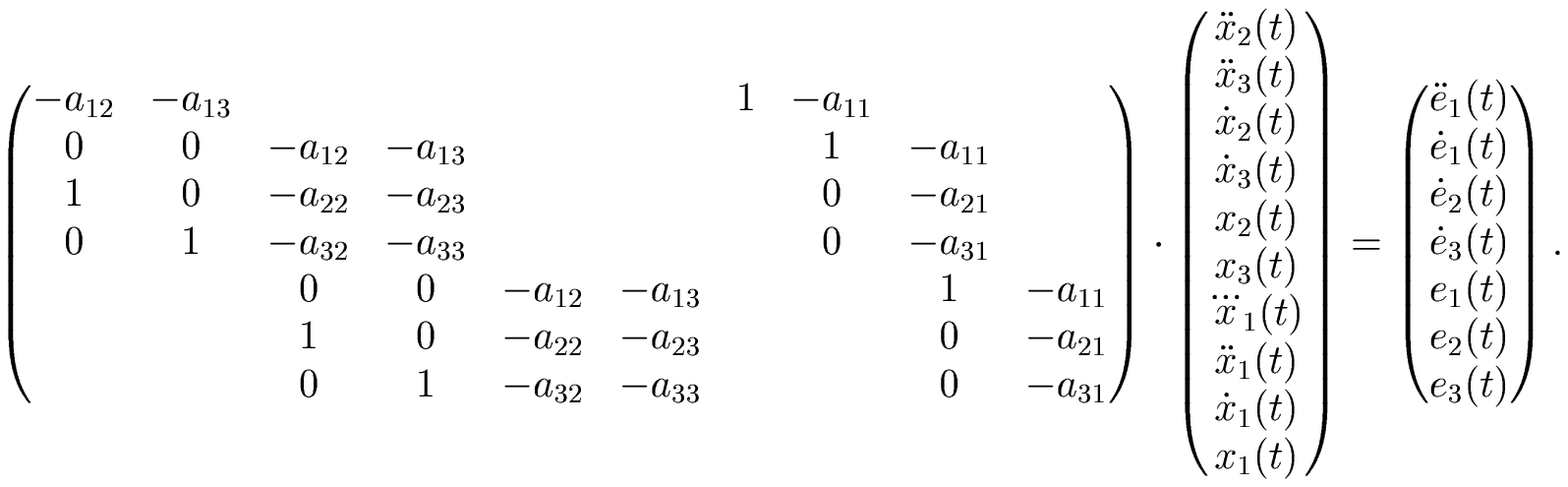}}
\vskip -0.5cm
\end{center}
\begin{equation}
\label{Bigsystem}\,
\end{equation}
%
\noindent
We point out the fact, that the variables should be arranged according to the {\sl ordering} 
$\ddot x_2(t) >$
$\ddot x_3(t)>$
$\dot x_2(t)>$
$\dot x_3(t)>$
$x_2(t)>$
$x_3(t)>$
$\dddot x_1(t)>$
$\ddot x_1(t)>$
$\dot x_1(t)>$
$x_1(t).$ 
For our algorithm to work, it is essential, that the last entries of the vector are the derivatives of the variable, for which we want to deduce the differential equation, in decreasing order. 


The final step is to use Gaussian elimination to convert the system into one in upper triangular form 
\begin{equation}
{\includegraphics[width=\linewidth,clip,keepaspectratio]{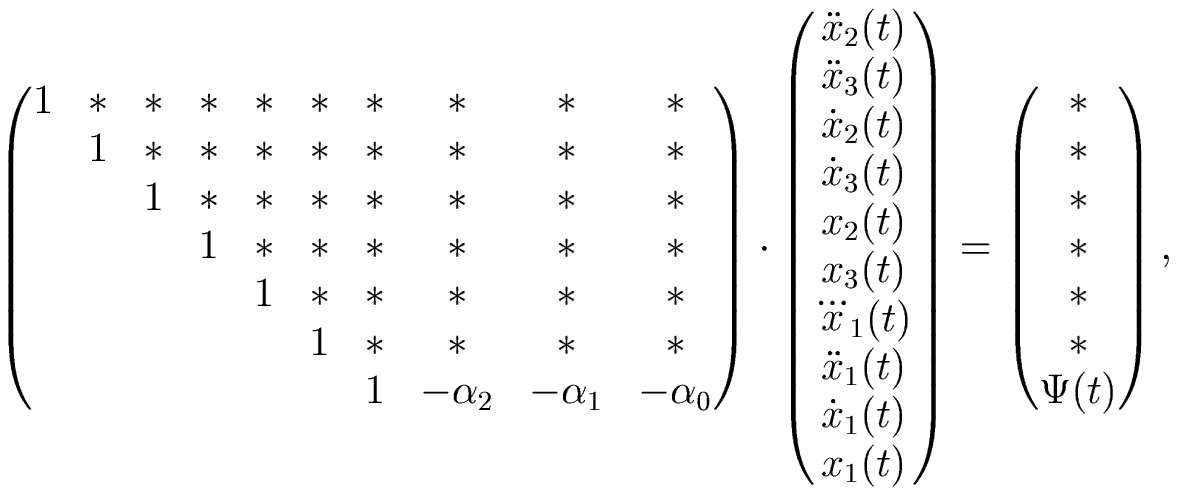}}
\label{Gaussian}
\end{equation}

\noindent
where $\Psi(t)$ is a linear combination of the functions
$\ddot e_1(t),$
$\dot e_1(t),$ $e_1(t),$
$\dot e_2(t),$ $e_2(t),$
$\dot e_3(t),$ and $e_3(t).$ 

Since we had $(n-1)\cdot n +1$ (obviously) linearly independent equations for $n^2+1$ variables, the last row will give one equation for the last $n+1$ variables, which according to our special ordering are all derivatives of $x_1$. Thus we have deduced the differential equation for $x_1$, which we were looking for. This is the same equation, which we would have found, if we had worked in the frequency domain and had calculated the transfer function for the Laplace transform ${\cal L}(x_1)$ of $x_1.$
\endproof

\subsection{Linear Semi\-State Equations}

It might happen, that a linear time-invariant circuit does not possess a description by state equations. Nevertheless it may be describable by so-called semi\-state equations, i.e.\ equations of the form
\begin{equation}
\label{semi}
{\bf E}{\bf \dot x}(t)  = {\bf A}{\bf x}(t) + {\bf B}{\bf u}(t),\quad
{\bf y}(t)  = {\bf C}{\bf x}(t) + {\bf D}{\bf u}(t),
\end{equation}
in which $\bf A,$ $\bf B,$ $\bf C,$ $\bf D$ and $\bf E$ are constant matrices, $\bf E$ being singular, ${\bf x}(t)$ denotes the semi\-state vector, ${\bf u}(t)$ the vector of inputs and ${\bf y}(t)$ the vector of outputs. 

The algorithm given above can be adapted to this situation, but we have to keep in mind, that even if $n$ semi\-state equations are given, the differential equation for any quantity might be of degree strictly less than $n$. 
We will see this effect in the example given below, which recently appeared in \citep{semistate}.

\smallskip
\begin{example}
{\bf (OTA-Circuit)}
\label{OTA}

\noindent
\begin{center}
{\includegraphics[width=\linewidth,height=7cm,clip,keepaspectratio]{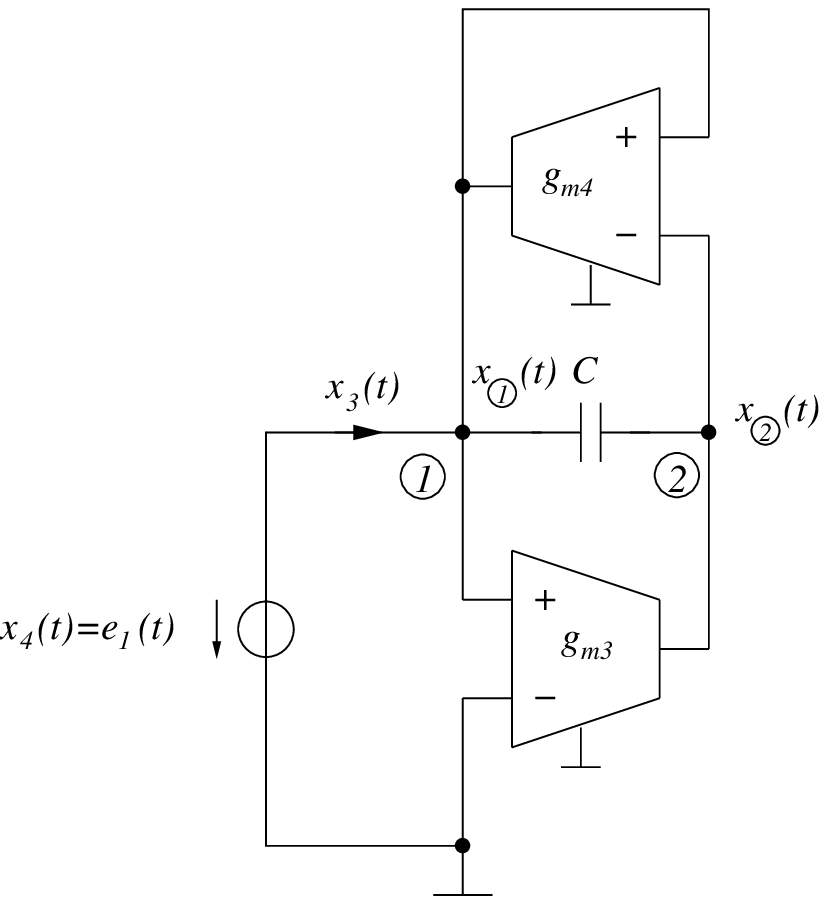}}
\bigskip
\end{center}
The OTA circuit shown above can be described by the semi\-state system
\begin{equation*}
\begin{split}
\begin{pmatrix}
C & -C & 0 \\
-C & C & 0 \\
0 & 0 & 0 
\end{pmatrix}
\cdot
\begin{pmatrix}
\dot x_{\xy*+{\scriptstyle{1}}*\cir{}\endxy}(t)\\
\dot x_{\xy*+{\scriptstyle{2}}*\cir{}\endxy}(t)\\
\dot x_3(t)
\end{pmatrix}
&=\\
=
\begin{pmatrix}
-g_{m4} & g_{m4} & 1 \\
-g_{m3} & 0 & 0 \\
1 & 0 & 0 
\end{pmatrix}
&\cdot
\begin{pmatrix}
x_{\xy*+{\scriptstyle{1}}*\cir{}\endxy}(t)\\
x_{\xy*+{\scriptstyle{2}}*\cir{}\endxy}(t)\\
x_3(t)
\end{pmatrix}
+
\begin{pmatrix}
0\\
0\\
-e_1(t)
\end{pmatrix},
\end{split}
\end{equation*}
where $x_{\xy*+{\scriptstyle{1}}*\cir{}\endxy}(t)$ and $x_{\xy*+{\scriptstyle{2}}*\cir{}\endxy}(t)$ denote the node to ground voltages of the respective nodes and $x_3(t)$ denotes the current into node ${\xy*+{1}*\cir{}\endxy}.$ 
%

As shown in \citep{semistate}, the transfer function 
$\displaystyle \underline{H}(s):= \frac{\underline{X_3}(s)}{\underline{E_1}(s)}$ is given by 
\begin{equation}\label{TransferOTA}\underline{H}(s) = g_{m3}\,\frac{Cs + g_{m4}}{Cs}.\end{equation}
We will deduce the corresponding differential equation for $x_3(t)$ in the time domain, using the above system of semi\-state equations and a slight modification of the algorithm for state equations.This algorithm is closer to the one, that will be used for nonlinear circuits. It is based on two main principles:

\smallskip
\begin{enumerate}
\item
The variables $x_{\xy*+{\scriptstyle{1}}*\cir{}\endxy},$ $x_{\xy*+{\scriptstyle{2}}*\cir{}\endxy},$ $x_3$ and their derivatives, are supposed to be ordered by
\begin{equation*}
\begin{split}
x_3
< \dot x_3
< \ddot x_3
< \dots
< 
x_{\xy*+{\scriptstyle{2}}*\cir{}\endxy}
< 
\dot x_{\xy*+{\scriptstyle{2}}*\cir{}\endxy}
< 
\ddot x_{\xy*+{\scriptstyle{2}}*\cir{}\endxy}
< \dots\\
< 
x_{\xy*+{\scriptstyle{1}}*\cir{}\endxy} 
< 
\dot x_{\xy*+{\scriptstyle{1}}*\cir{}\endxy} 
<
\ddot x_{\xy*+{\scriptstyle{1}}*\cir{}\endxy}
< \dots 
\end{split}
\end{equation*}
When a term is to be eliminated, we always choose the term of highest order.
\smallskip
\item
Equations are only differentiated when ''necessary''.
\end{enumerate}

\smallskip\noindent
So, let us get into details; the system was given by
\begin{alignat}{4}
\label{OTAa}
C \, &\dot x_{\xy*+{\scriptstyle{1}}*\cir{}\endxy}
- C \, &\dot x_{\xy*+{\scriptstyle{2}}*\cir{}\endxy}
& \, = \,& 
- g_{m4} \, x_{\xy*+{\scriptstyle{1}}*\cir{}\endxy}
& + g_{m4} \, x_{\xy*+{\scriptstyle{2}}*\cir{}\endxy} 
+ x_3\\
\label{OTAb}
- C \, &\dot x_{\xy*+{\scriptstyle{1}}*\cir{}\endxy}
+ C \, &\dot x_{\xy*+{\scriptstyle{2}}*\cir{}\endxy}
& \, = \,& 
- g_{m3} \, x_{\xy*+{\scriptstyle{1}}*\cir{}\endxy}
&\\
\label{OTAc}
&&0
& \, = \,& 
x_{\xy*+{\scriptstyle{1}}*\cir{}\endxy}
& - e_1
\end{alignat}
The term of highest order appearing in (\ref{OTAa}) - (\ref{OTAc}) is $x_{\xy*+{\scriptstyle{1}}*\cir{}\endxy}.$ Equation (\ref{OTAc}) gives
\begin{equation}
\label{x1elim}
x_{\xy*+{\scriptstyle{1}}*\cir{}\endxy} = e_1
\end{equation}
This can be used to eliminate $x_{\xy*+{\scriptstyle{1}}*\cir{}\endxy}$ from (\ref{OTAa}) and (\ref{OTAb}). Furthermore, after differentiating (\ref{x1elim}), we get
\begin{equation}
\label{DerX1elim} 
\dot x_{\xy*+{\scriptstyle{1}}*\cir{}\endxy} = \dot e_1.
\end{equation}
Thus we are able to remove all instances of $x_{\xy*+{\scriptstyle{1}}*\cir{}\endxy}$ in the first two equations. Consequently we are led to:
\begin{eqnarray}
\label{OTAd}
C\dot x_{\xy*+{\scriptstyle{2}}*\cir{}\endxy} & = & C\dot e_1 - g_{m3} e_1\\
\label{OTAe}
C\dot x_{\xy*+{\scriptstyle{2}}*\cir{}\endxy} & = & C\dot e_1 + g_{m4} e_1 - g_{m4} x_{\xy*+{\scriptstyle{2}}*\cir{}\endxy} - x_3
\end{eqnarray}
Clearly, these two equations imply the equality
\begin{equation}
\label{x2eq}
- g_{m3} e_1 = g_{m4} e_1 - g_{m4} x_{\xy*+{\scriptstyle{2}}*\cir{}\endxy} - x_3,
\end{equation}
which gives
\begin{equation}
\label{x2elim}
x_{\xy*+{\scriptstyle{2}}*\cir{}\endxy}= \frac{g_{m3}+g_{m4}}{g_{m4}} e_1 - \frac{x_3}{g_{m4}}
\end{equation}
and
\begin{equation}
\label{Derx2elim}
\dot x_{\xy*+{\scriptstyle{2}}*\cir{}\endxy}= \frac{g_{m3}+g_{m4}}{g_{m4}} \dot e_1 - \frac{\dot x_3}{g_{m4}}
\end{equation}
Putting this into (\ref{OTAe}), we arrive at
\begin{equation}
\label{kurzvorfinal}
C\dot e_1 - g_{m3}e_1=\frac{C}{g_{m4}}\cdot(g_{m3}+g_{m4})\dot e_1 - C\cdot \frac{\dot x_3}{g_{m4}},
\end{equation}
which finally results in
\begin{equation}
\label{Final}
C \cdot \dot x_3 = C\cdot g_{m3}\cdot \dot e_1 + g_{m3}\cdot{g_{m4}}\cdot e_1.
\end{equation}
This is the time domain equivalent of the transfer function (\ref{TransferOTA}), which we wanted to deduce.\endproof

\end{example}

\section{Constructive Differential Algebra and the {\tt diffalg}-package in {\tt maple}}

This is not the time and the space to give even a cursory treatment of those parts of differential algebra, which are needed to understand the sometimes subtle generalization to the nonlinear case of the algorithms shown above. For our purposes it is enough to know, that, cum grano salis, all the mechanisms are already visible in the example of the OTA circuit. 

Fortunately there already exist several implementations of the necessary algorithms within the computer algebra system {\tt MAPLE}. We will show by way of the above example the workings of one of these, the {\tt diffalg}-package, created by F.\ Boulier \citep{BLOP95}, \citep{BoulierDiss} and improved by E.\ Hubert \citep{HubertDiss} et al.\ A more detailed description (suitable for beginners with a mathematical background) can be found in the world wide web  \citep{HubertWebDiffAlg}, \citep{HubertWebOverview}.

\medskip
After starting a {\tt MAPLE}-session, one first has to load the package {\tt diffalg}:

\medskip
$
\tt > with(diffalg);
$

\smallskip
$
{[Rosenfeld\_Groebner, belongs\_to,delta\_leader,\dots ]}
\hfill\,
$

\bigskip
(The second line in slanted notation represents the output produced by {\tt diffalg}.)

After initialization, one has to enter the set of differential equations under consideration. This has to be done in form of so-called {\sl differential polynomials}. These are polynomials in the unknown functions $x_1,\dots,x_m,$ their (time) derivatives $x_i^{(\alpha)}:= D^{(\alpha)}x_i := \frac{d^\alpha}{dt^\alpha}x_i,$ $1\le i\le m,$ $\alpha\in \N$, the excitations $e_1,\dots, e_k$ and their derivatives, again.

In our example, we get three differential polynomials, which read in {\tt MAPLE}-notation:

\bigskip
$
\tt > p\_1 := C*diff(x\_1(t),t) - C*diff(x\_2(t),t)+g\_m4*x\_1(t)-g\_m4*x\_2(t)-x\_3(t);
$

$
\tt > p\_2 := - C*diff(x\_1(t),t) + C*diff(x\_2(t),t)+g\_m3*x\_1(t);
$

$
\tt > p\_3:= x\_1(t)-e\_1(t);
$

\bigskip
Next we have to tell the programme, which symbols it has to treat as constants. This is done with the command {\tt field\_extension}. {\tt diffalg} assumes, that we work over the rational numbers as ground field, where any further constants are considered as lying in a transcendental field extension of $\Q$ (i.e.\ we are allowed to divide by constants different from $0$, and constants do not satisfy any algebraic relations). If we work with symbols, e.g.\ for capacitors, resistors etc., this poses no problem. If we work with real coefficients (e.g.\ floating point numbers or algebraic numbers like $\sqrt 2$) major problems may arise. In our case we define

\bigskip
$
\tt > K := field\_extension(transcendental\_elements=[C,g\_m3,g\_m4]);
$

\smallskip
$
\hfill
{K := ground\_field}
\hfill\,
$

\bigskip
Finally we define a so-called {\sl differential ring}, which is supposed to contain all the objects of interest (i.e.\ differential polynomials and constants), and in which we are allowed to do the following operations
\begin{enumerate}
\item multiply a differential polynomials with a constant;
\item add and multiply  differential polynomials;
\item differentiate differential polynomials (if a constant is differentiated, the result is $0$).
\end{enumerate}

\smallskip
As we have seen above, it is very important, that we define an ordering on differential monomials. This is needed to control the elimination process. For this purpose, {\tt diffalg} asks for a ranking of the time dependent variables, from which it produces the obvious ''elimination ordering''. The variable, for which we want to know the differential equation, should be the last before the excitations.
\begin{equation*}
\begin{split}
\tt > R :=\, &\tt differential\_ring(ranking=
[x\_1,x\_2,x\_3,e\_1],\\ 
&\tt derivations=[t],
field\_of\_constants=K,\\
&\tt notation=diff);
\end{split}
\end{equation*}
$
\hfill
{R:=ODE\_ring}
\hfill\,
$

\medskip
The command {\tt Rosenfeld\_Groebner} lies at the heart of {\tt diffalg}. It produces minimal sets of differential polynomials generating the differential polynomials, we have entered.

$
\tt > GE := Rosenfeld\_Groebner(\{p\_1,p\_2,p\_3\}, R);
$

\smallskip
$
\hfill
{GE:=[regular]}
\hfill\,
$

{\tt GE} is a list and may contain several components\footnote{This is one of the subtleties of working with {\tt diffalg}, the details of which we do not want to present here.}. These components contain the sought-after differential equations, which can be listed with the help of the {\tt rewrite\_rules} command; in our example this gives 

\bigskip
$
\tt > rewrite\_rules(GE[1]);
$

\smallskip
$
{\left[x\_1(t)=e\_1(t), \right.}
$

\smallskip
$
\hfill
\displaystyle x\_2(t)=\frac{-x\_3(t)+e\_1(t)g\_m3+e\_1(t)g\_m4}{g\_m4},
\hfill\,
$

\smallskip
$
\hfill {\displaystyle\left.\frac{\partial}{\partial t}x\_3(t)=\frac{g\_m3\left(e\_1(t)g\_m4+C\left(\displaystyle\frac{\partial}{\partial t}e\_1(t)\right)\right)}{C}\right]}
$

\bigskip
Mark, that the last entry is the differential equation for $x_3,$ which we had deduced in example \ref{OTA}.


\section{Nonlinear Circuits and their corresponding differential equations}

Now, that we know, how the {\tt MAPLE} package {\tt diffalg} can be used to deduce  the time equivalent of the transfer function from any system of linear differential equations (with constant coefficients), in this section we will apply {\tt diffalg} in an analogous manner to several systems of nonlinear equations, which come from nonlinear circuits. 

\medskip
\begin{example}
{(\bf Damped resonant circuit with a nonlinear inductivity)}

As our first nonlinear example we have chosen a well known circuit from the literature, which leads to the Duffing equation (cp. \citep{Sternnonlinear}, example 11-1, or \citep{nonlinear}, chapter 1.3.2.).  

\smallskip\noindent
\begin{center}
{\includegraphics[width=\linewidth,height=4cm,clip,keepaspectratio]{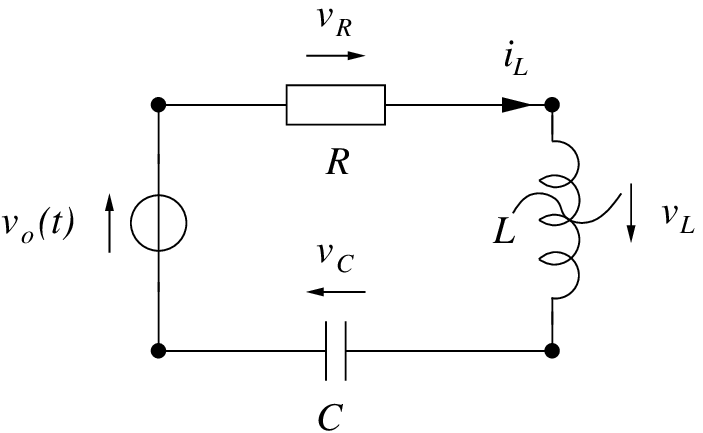}}
\end{center}
The resistor and the capacitor are assumed to be linear, i.e., they are described by
\begin{equation}
\label{DuffNW}
i_C(t)  =  C\dot v_C(t)\quad\hbox{and}\quad
v_R(t) = R\, i_R(t).
\end{equation}
The inductor is assumed to be nonlinear, being described by 
\begin{equation}
\label{DuffIL}
v_L(t) = \dot \Psi(t),
\end{equation}
where the current $i_L(t)$ is approximated by the cubic
\begin{equation}
\label{DuffCubic}
i_L(t) = a\cdot\psi(t) + b\cdot \psi(t)^3.
\end{equation}
Finally, Kirchhoff' s equation lead to
\begin{equation}
\label{DuffKirchhoff}
i_R(t) = i_L(t) = i_C(t)\ \hbox{and}\
v_R(t)+v_L(t)+v_C(t)=-v_0(t).
\end{equation}
Equations (\ref{DuffNW})-(\ref{DuffKirchhoff}) are translated into their corresponding differential polynomials and are used as input for the {\tt diffalg}-routine. This produces as part of the output of the subroutine {\tt rewrite\_rules}:
\begin{equation}
\begin{split}
\tt \frac{\partial^2}{\partial t^2}\psi(t)\,=\,
&-\frac{\displaystyle a\psi(t)+b\psi(t)^3+\left(\frac{\partial}{\partial t}v_0(t)\right)C}{C}\\
&-\frac{\displaystyle CRa\,\left(\frac{\partial}{\partial t}\psi(t)\right) + 3 CRb\,\psi(t)^2\left(\frac{\partial}{\partial t}\psi(t)\right)}{C},
\end{split}
\end{equation}
which is formula 1.30 in \citep{nonlinear} and is an equivalent of the Duffing equation:
\begin{equation}
\label{Duffing}
\frac{d^2}{dt^2}\psi = 
- (a + 3 b\cdot\psi^2)\cdot R \frac{d}{dt}\psi - \frac{a \psi}{C} - \frac{b \psi^3}{C}
-\frac{d}{dt}v_0(t) 
\end{equation}
\end{example}\endproof

\medskip\noindent
\noindent\begin{example}
{\bf (Chua's Circuit)}
Our next example is Chua's circuit, the first example of a physically realizable circuit, showing chaotic behavior \citep{ChuaGenesis}. This circuit consists of two linear capacitors, one linear inductor, one linear resistor and one nonlinear resistor, as shown below.

\medskip\noindent
\begin{center}
{\includegraphics[width=\linewidth,height=3.25cm,clip,keepaspectratio]{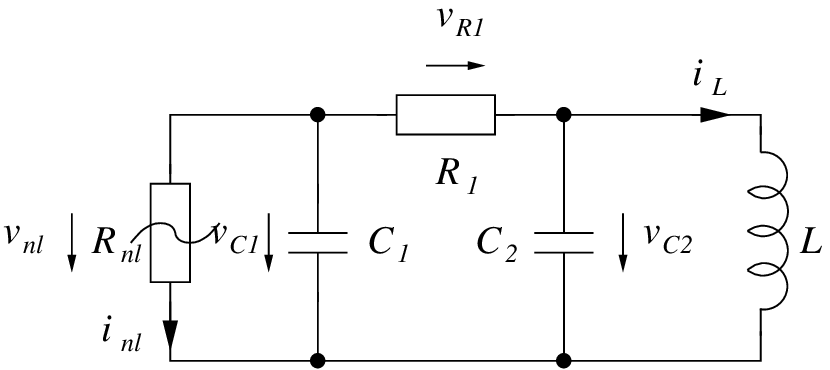}}
\end{center}
Since it is easy to write down the equations for the linear elements and the Kirchhoff equations, 
we concentrate on the mathematical description of the nonlinear resistor. The classic description of $R_{nl},$ using piecewise linear functions, is unsuitable for our purposes, because it does not satisfy the right differentiability conditions. Thus we use the one, presented e.g.\ in \citep{Chua_AI}, where the negative arctangent is used to produce the nonlinear $v$-$i$ characteristic of $R_{nl}.$ This leads to
\begin{equation}
\label{arctan}
i_{nl}(t) = - I_0\cdot \arctan\left(\frac{v_{nl}(t)}{V_0}\right).
\end{equation}
%
We are now confronted with a new problem: the arctangent is a transcendental function, thus (\ref{arctan}) would not give a differential polynomial, as needed. One procedural solution to this problem was presented before in \citep{Carra_SMACD}. 
In the present paper, we are satisfied with the fact, that (\ref{arctan}) implies yet another algebraic differential equation
\begin{equation}
\label{arctanDE}
\frac{d}{dt} i_{nl}(t) = -\frac{I_0}{V_0}\cdot\left(1 + \left(\displaystyle\frac{u_{nl}(t)}{V_0}\right)^2\right)^{-1}\cdot \frac{d}{dt} u_{nl}(t) 
\end{equation}
which obviously is given by a differential polynomial and which we can use as input for {\tt diffalg} instead of (\ref{arctan}).
From this equation together with the Kirchhoff equations and the characterizing equations of the linear elements, 
%
{\tt diffalg} produces the differential equation
\begin{equation}
\begin{split}
\label{ChuaSingle}
x^{(4)} =  -  \frac{1}{C_1 C_2 R L} \; &\cdot
\left( (C_1 + C_2) L\,\dddot x + C_1 R\, \ddot x + \dot x\phantom{\frac{V_0}{x^2 + V_0^2}}\right.\\
-&\, I_0\cdot \frac{V_0}{x^2 + V_0^2} \left[ C_2 L R\, \dddot x + L\, \ddot x + R\, \dot x\right]\\
 +&\,  2\, I_0 \cdot  \frac{V_0}{\left(x^2+V_0^2\right)^2} \cdot L\,x\,\dot x \cdot \left[ 3\, C_2 R\, \ddot x + \dot x \right]\\
-& \left. I_0\cdot \frac{V_0\, (6 x^2 - 2 V_0^2)}{\left( x^2 + V_0^2\right)^3}\; C_2 L R\, \dot x^3 
\right),
\end{split}
\end{equation}
where $x(t)$ denotes the voltage $v_{C_1}(t)$ through the capacitor $C_1.$ It has to be said that the final result given by {\tt diffalg} looks slightly different, since it is given in expanded form, i.e.\ the numerator consists of 31 summands. Formula (\ref{ChuaSingle}) has been reached at, only after some laborous post-processing. \endproof

%
\end{example}

\bigskip
\noindent
\begin{example}
\label{SimpleRectifier}
{\bf (Simple Model of a Peak Rectifier Circuit)} Now we are going to show how to handle diodes (and consequently by way of the Ebers-Moll model the large signal behaviour of BJTs) in electric circuits. As an example we have chosen a simple model of a peak rectifier circuit as seen in \citep{SedraSmith}, chapter 3.7 pp.\ 185ff.
\begin{center}
{\includegraphics[width=\linewidth,clip,keepaspectratio]{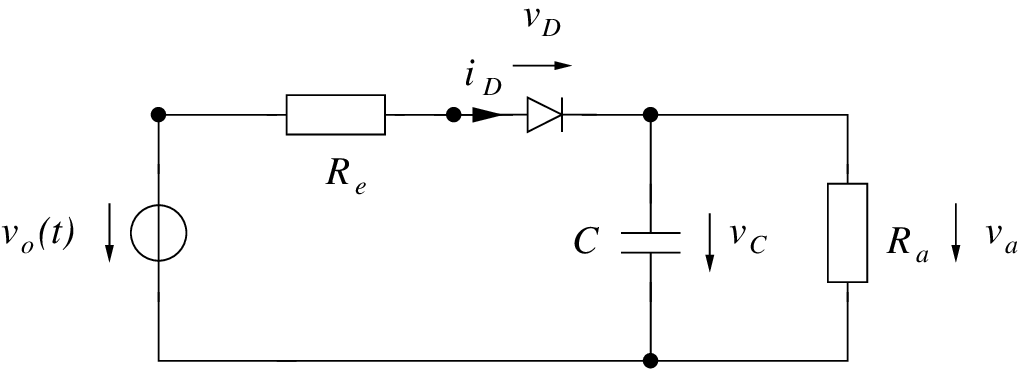}}
\end{center}
Again we concentrate on the only nonlinear element in the circuit -- the diode. It is well known, that the $v$-$i$ characteristic of a nonideal diode can be approximately described by 
\begin{equation}
\label{DiodeChar}
i_D(t) = I_s\cdot\left[ \exp\left(\frac{v_D(t)}{V_T}\right) - 1\right],
\end{equation}
where $I_s$ is the saturation current and $V_T$ is the thermal voltage -- 
quantities, which we consider constant during the course of our analysis.

As before we have to translate a transcendental equation into a differential polynomial. This can be done easily by differentiating (\ref{DiodeChar}) once, which due to the chain rule $\frac{d}{dt}\, i_D(t) = \frac{d}{dv_D}\, i_D \cdot \frac{d}{dt}\,v_D(t)$ results in the equation
\begin{equation}
\label{DiodeDiffpol}
\frac{d}{dt} i_D(t) = \frac{1}{V_T}\left(\frac{d}{dt}v_D(t)\right)\cdot [i_D(t) + I_s].
\end{equation}
%
%
%
%
This time {\tt diffalg} produces the following second order differential equation for the output voltage $v_a(t):$
\begin{equation}
\label{simpleRect}
\ddot v_a = - \frac{R_a\cdot [V_T \dot v_a - (\dot v_0 - \dot v_a)\cdot v_x]+ [ \dot v_a \cdot R_e v_x]}{C R_a\cdot \left(V_T R_a + R_e v_x\right)},
\end{equation}
where we have set 
$
v_x:= \left(C R_a \dot v_a + v_a + R_a I_s\right).
$

For the purpose of comparing this result to that appearing in the literature, we give the differential equation in case of an ideal voltage source, i.e.\ $R_e=0\Omega.$ It is given by
%
\begin{equation}
\label{SimpleRectIdealSource}
\ddot v_a
=
\frac{1}{C R_a}\left( - \dot v_a + \frac{1}{V_T} \left[\dot v_0-\dot v_a\right]\cdot
\left[C R_a\dot v_a + v_a + R_a I_s\right]\right).
\end{equation}
\end{example}\endproof

\section{Further Examples}

\begin{example}
{\bf (Diode Circuit with LC-Load and Nonideal Voltage source)}

\begin{center}
{\includegraphics[width=\linewidth,clip,keepaspectratio]{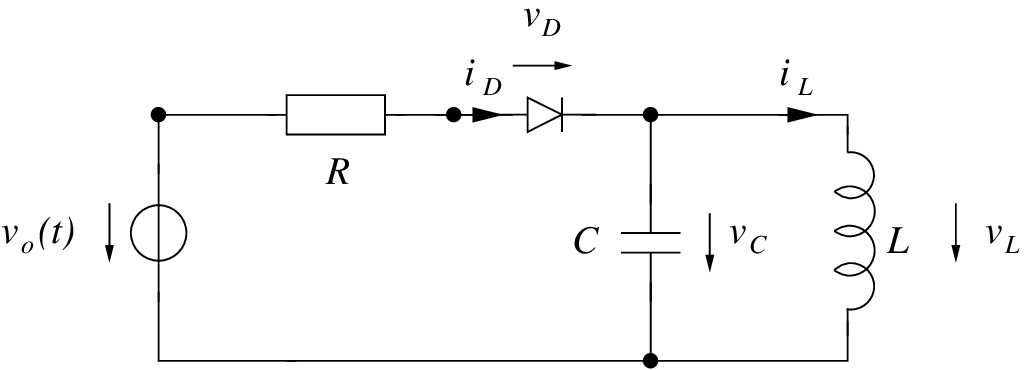}}
\end{center}

The above circuit may not be of much practical interest. Nevertheless it is a good test of the power of our approach (and the capabilities of {\tt diffalg}), since it slightly generalizes example \ref{SimpleRectifier} and we increase the number of dynamic elements in our circuit. Again, the diode is assumed to be nonideal, given by (\ref{DiodeDiffpol}). 

The output voltage $x(t) := v_C(t)$ is described by the equation
\begin{equation}
\label{LCDiode}
\begin{split}
\dddot x &=
- \left\{  \frac{1}{CL} \, \dot x
+ \frac{x + CL \ddot x}{\dot x - \dot v_0} \cdot \frac{1}{V_T\cdot CL}\cdot\right.\\ 
&\left.\left[ \left( \, \frac{R}{L}\, (x + CL \ddot x) + (\dot x - \dot v_0)\, \right)^2 - V_T\, (\ddot x - \ddot v_0) \right] \right\}.
\end{split}
\end{equation}

%
%
%
%
%
%
\end{example}\endproof

\bigskip
\noindent
In the course of our investigations, we have tried a number of larger circuits, which in principle are accessible to our approach. We met two main obstacles, which are natural in the ''business'' of symbolic methods:
\begin{enumerate}
\item 
the combinatorial explosion, resulting in a larger and larger number of terms contained in the final differential equation, and
\item the massive increase in time, needed by {\tt diffalg} to produce this equation. 
\end{enumerate}

\medskip\noindent
In the sequel we give a short report on these experiments:

\medskip
\begin{example}
{\bf (Peak Rectifier with Power Transformer)}
\smallskip

\hskip -1.2cm
\begin{minipage}{\linewidth}
\begin{center}
{\includegraphics[width=1.125\linewidth,clip,keepaspectratio]{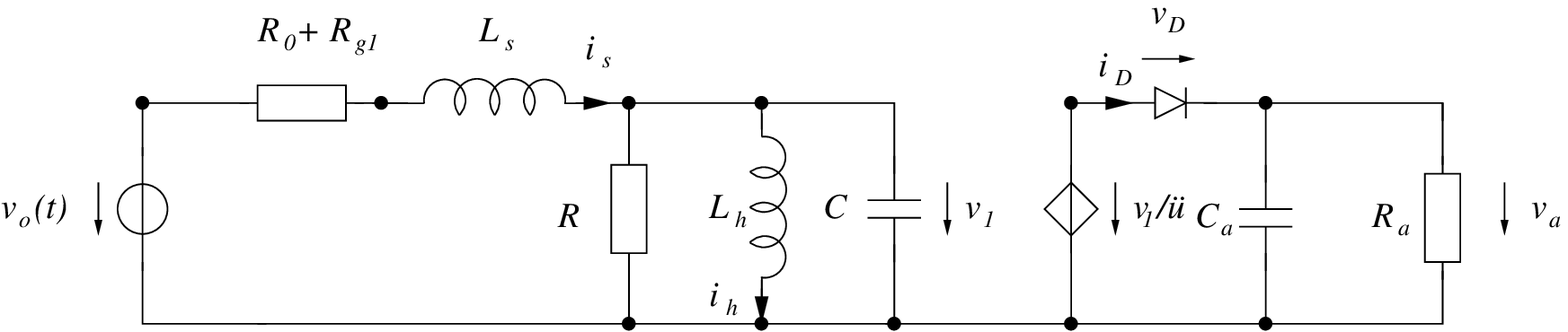}}
\end{center}
\end{minipage}

\medskip
This kind of circuit is described in the introduction to chapter 3.7 in \citep{SedraSmith}. As shown above, we have used the model given by Horneber in his PhD-thesis \citep{HorneberDiss}, section 14.1. There, it is the smaller of two examples, the other being the ''Ring Modulator'', which has become a benchmark in the numerical analysis of initial value problems \citep{IVTestSet}.

From our point of view, we can tell, that {\tt diffalg}, although needing substantial more time  than in the previous examples (1-2 minutes instead of only seconds), is able to produce a fifth order differential equation for the output voltage $v_a$. Unfortunately, we are not able to reproduce this result here, since the initial output even after some simplifications contains more than 600 summands. Thus some more ''post processing'' is needed to get an intelligible result. Even with the help of other facilities of {\tt MAPLE} this work has not been finished, yet\footnote{Note added in 2008: Meanwhile these calculations have been done. The end result still is to unwieldy to be presented here. Furthermore the collecting and combining of fully symbolic terms by hand has turned out to be so error-prone that, in the opinion of the author, some kind of additional ``plausibility measures'' need to be introduced.}.\endproof
\end{example}

\bigskip\noindent
Finally we have tried our approach on a {\bf ''simple'' single-stage common-emitter amplifier} (\citep{SedraSmith}, chapter 4.11) as modelled in \citep{SommerSymbolic} and on the above mentioned {\bf ring modulator} of Horneber. In both cases, up until now, even though we have used several days of computing time, we were not able to produce any results. Although the latter -- very ambitious -- example (which presumably will lead to an differential equation of order 18) may be beyond the scope of any computer algebra system for some time, the first should be within our grasp and should be attacked further. 

\section{Conclusion}

In this paper we have shown, how, using constructive methods from differential algebra and one of their \hbox{realizations --} the package {\tt diffalg} of the computer algebra system \hbox{{\tt MAPLE} --} linear and nonlinear circuits can be described by a single differential equation. 
In the future it will be necessary to further examine the power of this approach, i.e.\ to find more and larger circuits, which can be treated this way. Furthermore, if the number of these circuits is large enough, methods have to be found, that allow a fast and ''easy'' analysis of the resulting equations, analogous to the analysis of linear circuits by way of their transfer functions.
 
\bibliographystyle{IEEE}

\section*{Note added to the Electronic Version}
In this electronic document, some small typographical errors of the printed version were corrected. This especially refers to formulas (\ref{kurzvorfinal}) and (\ref{Final}).

Furthermore, for the convenience of the reader the abstract has been rewritten, and keywords, an MSC classification, and a short CV according to IEEE standards have been added. URLs have been checked again, and, where necessary, have been updated.  Finally the dedication has been expanded. The main body of the article, however, remains unchanged.\phantom{mmmmmmmmm}\hfill 
\phantom{m}\hfill (April~17th,~2008)

\begin{biography}
[{\includegraphics[width=1in,height=1.25in,clip,keepaspectratio]{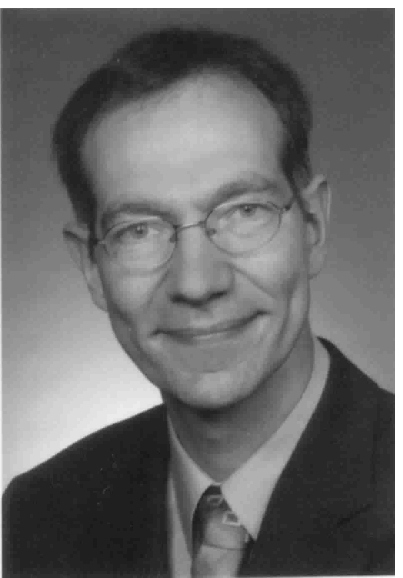}}]
{Eberhard H.-A.~Gerbracht}
received a Dipl.-Math.\ degree in mathematics, a Dipl.-Inform.\ degree in computer science, and a Ph.D. (Dr.\ rer.nat.) degree in mathematics from the Technical University Braunschweig, Germany, in 1990, 1993, and 1998, respectively.

From 1992 to 1997 he was a Research Fellow and Teaching Assistant at the Institute for Geometry at the TU Braunschweig. From 1997 to 2003 he was an Assistant Professor in the Department of Electrical Engineering and Information Technology at the TU Braunschweig. During that time he was also appointed lecturer for several courses on digital circuit design at the University of Applied Sciences Braunschweig/Wolfenb\"uttel, Germany. From 2001 to 2002 he was appointed lecturer for a two-semester course in linear circuit analysis at the TU Braunschweig. After a two-year stint as a mathematics and computer science teacher at a grammar school in Braunschweig and a vocational school in Gifhorn, Germany, he is currently working as advisor, and independent researcher in various areas of mathematics. His research interests include combinatorial algebra, C*-algebras, the history of mathematics in the 19th and early 20th century and applications of computer algebra and dynamical geometry to graph theory, calculus, and electrical engineering.

Dr.~Gerbracht is a member of the German Mathematical Society (DMV), the German Society for Didactics of Mathematics (GDM), and founding member of the society ``Web Portal: History in Braunschweig - \href{http://www.gibs.info}{www.gibs.info}''.
\end{biography}

\end{document}